\input harvmac

\noblackbox

\def\tilde{\widetilde}
\newcount\figno
\figno=0
\def\fig#1#2#3{
\par\begingroup\parindent=0pt\leftskip=1cm\rightskip=1cm\parindent=0pt
\baselineskip=11pt
\global\advance\figno by 1
\midinsert
\epsfxsize=#3
\centerline{\epsfbox{#2}}
\vskip 12pt
\centerline{{\bf Figure \the\figno:} #1}\par
\endinsert\endgroup\par}
\def\figlabel#1{\xdef#1{\the\figno}}

\def\pl#1#2#3{Phys. Lett. {\bf B#1} (#2) #3}

\def\physrev#1#2#3{Phys. Rev. {\bf D#1} (#2) #3}
\def\ap#1#2#3{Ann. Phys. {\bf #1} (#2) #3}

\def\cqg#1#2#3{Class. Quant. Grav. {\bf #1} (#2) #3}


\font\cmss=cmss10
\font\cmsss=cmss10 at 7pt
\def\rlx{\relax\leavevmode}
\def\inbar{\vrule height1.5ex width.4pt depth0pt}
\def\IC{\relax\,\hbox{$\inbar\kern-.3em{\rm C}$}}
\def\IN{\relax{\rm I\kern-.18em N}}
\def\IP{\relax{\rm I\kern-.18em P}}
\def\ZZ{\rlx\leavevmode\ifmmode\mathchoice{\hbox{\cmss Z\kern-.4em Z}}
 {\hbox{\cmss Z\kern-.4em Z}}{\lower.9pt\hbox{\cmsss Z\kern-.36em Z}}
 {\lower1.2pt\hbox{\cmsss Z\kern-.36em Z}}\else{\cmss Z\kern-.4em
 Z}\fi} 
\def\IZ{\relax\ifmmode\mathchoice
{\hbox{\cmss Z\kern-.4em Z}}{\hbox{\cmss Z\kern-.4em Z}}
{\lower.9pt\hbox{\cmsss Z\kern-.4em Z}}
{\lower1.2pt\hbox{\cmsss Z\kern-.4em Z}}\else{\cmss Z\kern-.4em
Z}\fi}

\def\narrowplus{\kern -.04truein + \kern -.03truein}
\def\narrowminus{- \kern -.04truein}
\def\narrowminussub{\kern -.02truein - \kern -.01truein}

\def\SYMF{SYM$_4$}

\def\half{{1\over 2}}

\def\m{{\mu}}
\def\n{{\nu}}

\def\b{{\beta}}
\def\a{{\alpha}}
\def\ga{{\gamma}}
\def\ep{{\epsilon}}
\def\d{{\delta}}
\def\G{{\Gamma}}
\def\ph{{\phi}}

\def\l{{\lambda}}

\def\s{{\sigma}}

\def\del{{\nabla}}

\def\CC{{\cal C}}

\def\CN{{\cal N}}

\def\CO{{\cal O}}

\def\CN{{\cal N}}

\def\ca{\langle \CC^{I_1} \CC^{I_2} \rangle}

\def\cabc{\langle \CC^{I_1} \CC^{I_2} \CC^{I_3} \rangle}

\def\IZ{\relax\ifmmode\mathchoice
{\hbox{\cmss Z\kern-.4em Z}}{\hbox{\cmss Z\kern-.4em Z}}
{\lower.9pt\hbox{\cmsss Z\kern-.4em Z}}
{\lower1.2pt\hbox{\cmsss Z\kern-.4em Z}}\else{\cmss Z\kern-.4em
Z}\fi}
\def\IB{\relax{\rm I\kern-.18em B}}
\def\IC{{\relax\hbox{$\inbar\kern-.3em{\rm C}$}}}
\def\ID{\relax{\rm I\kern-.18em D}}
\def\IE{\relax{\rm I\kern-.18em E}}
\def\IF{\relax{\rm I\kern-.18em F}}
\def\IG{\relax\hbox{$\inbar\kern-.3em{\rm G}$}}
\def\IGa{\relax\hbox{${\rm I}\kern-.18em\Gamma$}}
\def\IH{\relax{\rm I\kern-.18em H}}

\def\II{\relax{\rm I\kern-.18em I}}
\def\IK{\relax{\rm I\kern-.18em K}}
\def\IP{\relax{\rm I\kern-.18em P}}

\def\p{\partial}

\font\cmss=cmss10 \font\cmsss=cmss10 at 7pt
\def\IR{\relax{\rm I\kern-.18em R}}

\def\e{{\epsilon_0}}

%

%
%
\def\eqnn#1{\xdef #1{(\secsym\the\meqno)}\writedef{#1\leftbracket#1}%
\global\advance\meqno by1\wrlabeL#1}
\def\eqna#1{\xdef #1##1{\hbox{$(\secsym\the\meqno##1)$}}
\writedef{#1\numbersign1\leftbracket#1{\numbersign1}}%
\global\advance\meqno by1\wrlabeL{#1$\{\}$}}
\def\eqn#1#2{\xdef #1{(\secsym\the\meqno)}\writedef{#1\leftbracket#1}%
\global\advance\meqno by1$$#2\eqno#1\eqlabeL#1$$}

%
%

\lref\rmal{J. Maldacena, 
``The Large $N$ limit of Superconformal Field Theories and Supergravity,'' 
hep-th/9711200.}
\lref\rwit{E. Witten, ``Anti-de Sitter Space and Holography,''
hep-th/9802150.}
\lref\rpol{S. S. Gubser, I. R. Klebanov and A. M. Polyakov, ``Gauge Theory
Correlators from Non-critical String Theory,'' hep-th/9802109.}

\lref\rvishvanath{W. M\"uck and  K. S. Viswanathan, ``Conformal Field
Theory Correlators from Classical Scalar Field Theory on $AdS_{d+1}$,''
hep-th/9804035.}
\lref\rvishvanathb{W. M\"uck and  K. S. Viswanathan, ``Conformal Field
Theory Correlators from Classical Field Theory on Anti-de Sitter Space II.
Vector and Spinor Fields,'' hep-th/9805145.}
\lref\rmathur{D. Z. Freedman, S. D. Mathur, A. Matusis and L. Rastelli, 
``Correlation Functions in the CFT$_d$/$AdS_{d+1}$ Correspondence,'' 
hep-th/9804058.}
\lref\ramir{A. M. Ghezelbash, K. Kaviani, S. Parvizi and  A. H. Fatollahi,
``Interacting Spinors-Scalars and the AdS/CFT Correspondence,''
hep-th/9805162.}
\lref\rtsytlin{H. Liu and A. A. Tseytlin, ``D=4 Super-Yang-Mills, D=5 Gauged
Supergravity, and D=4 Conformal Supergravity,'' hep-th/9804083.}

\lref\rdas{S. R. Das and S. P. Trivedi, ``Three Brane Action and the
Correspondence between $\CN=4$ Yang-Mills Theory and Anti-de Sitter Space,''
hep-th/9804149.}
\lref\rzaf{S. Ferrara, M. A. Lledo, A. Zaffaroni, ``Born-Infeld Corrections
to the D3-brane action in ADS(5) $\times$ S(5) and $\CN=4$, $D=4$ Primary
Superfields,'' hep-th/9805082.}

\lref\rkim{H. J. Kim, L. J. Romans and P. van Nieuwenhuizen 
``Mass Spectrum of Chiral Ten-dimensional $\CN=2$ Supergravity on $S^5$,''
\physrev{32}{1985}{389}.}
\lref\rgunaydin{M. Gu\"naydin and N. Marcus, ``The Spectrum of the $S^5$ 
Compactification of the Chiral $\CN=2$, $D=10$ Supergravity and the Unitary Supermultiplets of $U(2,2|4)$,'' \cqg{2}{1985}{L11-L17}.}

\lref\rbanks{T. Banks and M. B. Green, ``Non-perturbative Effects in 
$AdS_5 \times S^5$ String Theory and d=4 SUSY Yang-Mills,''
hep-th/9804170}

\lref\rtsorkin{G. Dall'Agata, K. Lechner and D. Sorokin, 
``Covariant Actions for the Bosonic Sector of D=10 IIB supergravity,'' 
hep-th/9707044 and references therein.}
\lref\rchodos{A. Chodos and E. Myers, 
``Gravitational Contribution to the Casimir Energy in Kaluza-Klein 
Theories,''
\ap{156}{1984}{412-441}.}
\lref\rmtw{C. W. Misner, K. S. Thorn, J. A. Wheeler, ``Gravitation,'' 
W. H. Freeman and Company (1973), p. 965.}
\lref\rminwalla{S. Minwalla, 
``Restrictions imposed by Superconformal Invariance On 
Quantum Field Theories,
''hep-th/9712074.}
\lref\rdobrev{V. K. Dobrev and V. B. Petkova, ``All Positive 
Energy Unitary 
Irreducible Representations Of Extended Conformal Supersymmetry,''
\pl{162}{1985}{127-132}.} 

\lref\rchalmers{G. Chalmers, H. Nastase, K. Schalm and R. Siebelink
 ``$R$-Current Correlators in $\CN=4$ SYM from $AdS$,''
hep-th/9805015.}
\lref\rnonren{P. S. Howe and P. C. West, ``Is $\CN=4$ Yang-Mills Theory 
Soluble?'' hep-th/9611074.}
\lref\rosborn{J. Erdmenger and H. Osborn, 
``Conserved Currents and the Energy Momentum Tensor 
in Conformally Invariant Theories for General Dimensions,'' hep-th/9605009.}
\lref\rklebone{S. S. Gubser, I. R. Klebanov, and A. W. Peet,
``Entropy And Temperature Of Black 3-Branes,'' Phys. Rev. {\bf D54}
(1996) 3915, hep-th/9602135.} 
\lref\rklebtwo{I. R. Klebanov, ``World Volume Approach To Absorption By
Nondilatonic Branes,'' Nucl. Phys. {\bf B496} (1997) 231, hep-th/9702077.}
\lref\rklebthree{S. S. Gubser, I. R. Klebanov, A. A. Tseytlin,
``String Theory And Classical Absorption By Three-branes,''
Nucl. Phys. {\bf B499} (1997) 217, hep-th/9703040.}
\lref\rklebfour{S. S. Gubser and I. R. Klebanov, ``Absorption By Branes And
Schwinger Terms In The World Volume Theory,'' Phys. Lett. {\bf B413}
(1997) 41, hep-th/9708005.} 

\lref\rsfetsos{M.Henningson and K. Sfetsos, ``Spinors and the 
$AdS/CFT$ correspondence'', hep-th/9803251}
\lref\rzaf{S.Ferrara and A Zaffaroni,``On N=8 Supergravity on 
$AdS_5$ and $N=4$ Superconformal Yang-Mills therory'', hep-th/9802203 .}

\Title
{\vbox{
\baselineskip12pt
\hbox{hep-th/9806074}\hbox{PUPT-1796}\hbox{IASSNS-HEP-98-51}
}}
{\vbox{
\centerline{Three-Point Functions of Chiral Operators}
\centerline{in $D=4$, $\CN=4$ SYM at Large $N$}
}}

\centerline{
Sangmin Lee, Shiraz Minwalla, Mukund Rangamani
\footnote{$^1$}{sangmin, minwalla, rangamni@princeton.edu}
}
\smallskip
\centerline{\sl Department of Physics, Princeton University}
\centerline{\sl Princeton, NJ 08544, USA}
\medskip
\centerline{and}
\medskip
\centerline{Nathan Seiberg\footnote{$^2$}{seiberg@sns.ias.edu}}
\smallskip
\centerline{\sl School of Natural Sciences, Institute for Advanced Study}
\centerline{\sl Olden Lane, Princeton, NJ 08540, USA}

\vskip 0.8cm

\centerline{\bf Abstract}
\medskip
\noindent 
We study all three-point functions of normalized chiral operators in
$D=4$, $\CN=4$, $U(N)$ supersymmetric Yang-Mills theory in the large
$N$ limit.  We compute them for small 't Hooft coupling $\lambda
=g_{YM}^2N \ll 1$ using free field theory and at strong coupling
$\lambda =g_{YM}^2N \gg 1$ using the $AdS$/CFT correspondence.
Surprisingly, we find the same answers in the two limits.  We
conjecture that at least for large $N$ the exact answers are
independent of $\lambda$.

\vskip 0.5cm
\Date{June 1998}


\newsec{Introduction}


The conjectured duality \rmal\ (for earlier related references see
\refs{\rklebone, \rklebtwo, \rklebthree, \rklebfour}) 
between string/M theory on Anti-de
Sitter space ($AdS$) times a compact manifold, and conformal field
theory (CFT) living on the boundary of $AdS$ has attracted much
attention. According to this proposal, Type IIB string theory on
$AdS_5\times S^5$ is dual to $D=4$, $\CN=4$ supersymmetric Yang-Mills
theory (\SYMF).

In \refs{\rpol, \rwit} a detailed dictionary relating S-matrix
elements of the string theory to Green's functions of the CFT was
proposed.  The operators of the CFT are mapped to on shell bulk fields
on $AdS$. The CFT operators interact with the boundary values of
these bulk fields through an interaction action $S_{int}$. The
partition function of the string theory with fixed boundary values of
fields is then identified with the partition function of the CFT with
external sources coupled to the corresponding operators.

Using this dictionary, two point functions of CFT operators 
corresponding to massive scalars
\refs{\rpol , \rwit , \rvishvanath, \rmathur}, vectors \refs{\rwit, 
\rmathur }, the graviton \rtsytlin, and spinors \rsfetsos\ have been 
computed. 

In a series of recent papers, the 3-point functions of operators in a
CFT$_4$ corresponding to massive minimally coupled
scalars \refs{\rvishvanath, \rmathur}, or scalars and spinors \ramir,
or vectors and spinors \rvishvanathb\ on the $AdS_5$ with certain
generic, arbitrarily prescribed, interactions have been computed.

Certain computations of correlation functions of operators in actual
\SYMF\ have also been performed.  Using a proposed form of $S_{int}$,
the 2-point functions of the stress energy tensor and ${\bf
Tr}(F_{\m\n}F^{\m\n})$ were computed in \rpol.  Using the model
independent coupling of gauge fields to currents, the 3-point
functions of the $R$-symmetry currents of \SYMF\ were computed in
\refs{\rmathur , \rchalmers}. Similarly, 3-point functions of the
dilaton and the stress energy tensor were computed in \rtsytlin.

Local operators in \SYMF\ are organized into infinite dimensional
families, each of which is an irreducible representation of the $D=4$,
$\CN=4$, superconformal algebra. Each family (or module) contains
special operators of lowest scaling dimension in an $SU(4)$
representation.   We will call them  primary operators (PO) (strictly,
only the operator with the highest $SU(4)$ weight is primary).
\SYMF\ contains a set of special short families that contain fewer
operators than the normal module. Such families include primary
operators which are chiral under an $\CN=1$ subalgebra; the scaling
dimension of operators in these families is determined by the
superconformal algebra \refs{\rdobrev, \rminwalla}\ in terms of their
$SU(4)$ $R$-symmetry representation.  We will loosely refer to all the
lowest dimension operators in such a representation as chiral primary
operators (CPO).  Under a given $\CN=1$ subalgebra, the $\CN=4$ chiral
primaries include $\CN=1$ chiral operators, $\CN=1$ anti-chiral
operators and non-chiral operators.

It should be stressed that unlike the situation in $\CN=1$, these
chiral primary fields do not form a ring.  The product of two $\CN=4$
chiral operators includes a product of an $\CN=1$ chiral operator with
an $\CN=1$ anti-chiral operator and even two $\CN=1$ non-chiral
operators, which are singular.  Because of such singularities the
$\CN=4$ chiral operators do not form a ring.

In this paper, we study the 3-point functions of all CPOs, in the
large $N$ limit of \SYMF.  We first compute them in the limit of weak
't Hooft coupling $\lambda = g_{YM}^2N\ll 1$ using free field theory.
We then study them in the limit of large 't Hooft coupling $\lambda =
g_{YM}^2N\gg 1$ using Type IIB supergravity (SUGRA).  Surprisingly, we find
the same answers.  Clearly, this agreement for the primary fields
guarantees similar agreements for all their descendants.

Banks and Green \rbanks\ showed that for infinite $N$ the leading order
result at large $\lambda$ is not corrected at the next order.  Given
that we found that the leading order result agrees with the weak
coupling answer, we are led to conjecture that {\it the 3-point
functions of all chiral primary operators at large $N$ is independent
of $\lambda = g_{YM}^2N$.}  

Since $R$-symmetry currents and the stress energy tensor are
descendents of CPOs, our results include all previous results on
3-point functions \refs{\rmathur, \rtsytlin , \rchalmers }, as
special cases. Also, the discussion of \refs{\rosborn, \rnonren} 
shows that some of these 3-point functions are independent of the
coupling even for finite $N$.

We point out that a similar result cannot be true for the
4-point function of these chiral operators.  Unlike the 3-point
functions, the 4-point functions depend on $\lambda$ at the next to
leading order \rbanks.

It might be that even a stronger claim is true, and these 3-point
functions are independent of $g_{YM}$ even for finite $N$.  (For some
of the 3-point functions this was proven in \rnonren.)  From the
weak coupling side it is clear that the 3-point functions depend
on $N$ (even the spectrum of chiral primary operators depends on $N$).
Therefore, if this stronger claim is true, then at strong coupling, on
the $AdS$ side, the coupling of three gravitons depends on $N$; i.e.\
it is corrected by quantum stringy effects.  It is well known that
such corrections are absent around flat space.  This result is a
consequence of the large amount of supersymmetry in the flat space
theory. Since the $AdS_5\times S^5$ background preserves the same
number of supersymmetries as the flat space background, one might
guess that here too the scattering of three gravitons is not affected
by quantum corrections. This guess cannot be simultaneously correct
with the claim that the 3-point functions are not corrected at
finite\foot{We thank T. Banks for a useful discussion on this
point.} $N$.

This paper is organized as follows.  In section 2, we compute the
correlation functions in the weak coupling limit. In section 3, we
identify fields on $AdS$ which represent the modes corresponding to
the chiral operators and construct their effective action to cubic
order in the fields.  In section 4, we use this action to obtain the
3-point functions of normalized CPOs of the SYM$_4$.  We compare this
result with the free field calculation of section 2 and find precise
agreement.  In Appendix A, we explain our notations and conventions.
Appendix B is devoted to spherical harmonics on $S^5$; we define
scalar, vector and tensor spherical harmonics in arbitrary dimensions,
and obtain several formulae needed for the calculation in section 3.

\newsec{Correlation Functions at weak coupling}


CPOs of \SYMF\ are operators of the form
$$\CO^I = \CC^I_{i_1..i_k}{\bf Tr}(\ph^{i_1}..\ph^{i_k}),$$ 
where $i_1, \cdots ,i_k$ are $SO(6)$ vector indices and
$\ph^i$ are six  $N \times N$ matrices transforming 
in the adjoint of $U(N)$. The trace in the formula above is 
over $U(N)$ indices.
$\CC^I$ is a totally symmetric traceless rank $k$ tensor of
$SO(6)$. We can choose an orthonormal basis on the vector space
$\{\CC^I\}$ such that $\ca \equiv \CC^{I_1}_{i_1 \cdots i_k}
{\CC^{I_2}}^{i_1 \cdots i_k} = \delta^{I_1I_2}$. 
We normalize our action as 
$\  \ S  = - \int{1 \over 2 g_{YM}^2} {\bf Tr} F^2 + \cdots  
= - \int {1 \over 4 g_{YM}^2} F_{\m\n}^a F^{a\m\n} + \cdots.$ 
In this normalization the Yang-Mills coupling and the string coupling
are  related by $g_{YM}^2 = 4 \pi g_s$.  The propagators of interest are
$$\langle\ph^i_a(x)\ph^j_b(y)\rangle=
{g_{YM}^2 \d_{ab} \d^{ij} \over (2 \pi)^2 |x-y|^2},$$
where $a,b, \ldots$ are $U(N)$ color indices.

The 2-point function of two CPOs specified by tensors
$\CC^{I_1}_{i_1...i_{k_1}}$ and $\CC^{I_2}_{j_1...j_{k_2}}$, is
computed in free field theory by contracting all the $\ph$s pair-wise
and is nonzero only if $k_1=k_2=k$.  Consider
$$g(x,y)=\langle{\bf Tr}(\ph^{i_1}(x)..\ph^{i_k}(x)) 
{\bf Tr}(\ph^{j_1}(y)...\ph^{j_k}(y))\rangle.$$
In the large $N$ limit only planar diagrams contribute. Planar diagrams 
correspond to contracting $i's$ and $j's$ in the same cyclic order in which 
they appear in $g(x,y)$. One finds
$$g(x,y)={N^k g_{YM}^{2k}(\d^{i_1j_1}\d^{i_2j_2}..\d^{i_kj_k} + {\rm cyclic})
\over (2 \pi)^{2k} |x-y|^{2k}}.$$
Using the orthonormality of the $\CC$ coefficients one thus deduces that
(the $\d^{I_1I_2}$ term in the equation below is replaced by 
$\ca $ when considering the 2-point function of arbitrary CPOs which
are not necessarily orthogonal)
\eqn\twopt{
\langle \CO^{I_1}(x) \CO^{I_2}(y) \rangle = 
\l^k {k \over (2 \pi)^{2k} |x-y|^{2k}} \delta^{I_1I_2}.
}
  
In a similar fashion one may compute the 3-point function of CPOs 
specified by $\CC^{I_1}_{i_1..i_{k_1}}, \CC^{I_2}_{j_1..j_{k_2}} , 
\CC^{I_3}_{l_1..l_{k_3}}$. To ensure that all $\ph$s are contracted, 
$\a_3={k_1+k_2-k_3 \over 2}$ of the $\ph$s must contract between the
first and second of these operators and similarly for other pairs.  In
the large $N$ limit, one finds
\eqn\threept{
\langle \CO^{I_1} \CO^{I_2} \CO^{I_3} \rangle
= {\l^{\Sigma/2} \over N} {k_1k_2k_3 \over (2 \pi)^{\Sigma}
|x-y|^{2\a_3} |y-z|^{2\a_1} |z-x|^{2\a_2}} \cabc,
}
where $\Sigma = k_1 +k_2 + k_3$ and $\cabc$ represents the unique
$SO(6)$ invariant that can be formed from $\CC^{I_1}, \CC^{I_2},
\CC^{I_3}$ (by contracting $\a_1$ indices between $\CC^{I_2}$ and $\CC^{I_3}$;
$\a_2$ indices between $\CC^{I_3}$ and $\CC^{I_1}$ and 
$\a_3$ indices between $\CC^{I_1}$ and $\CC^{I_2}$)

We rescale the CPOs $O^{I}=\CO^I{(2\pi)^k \over \l^{k/2} \sqrt{k}}$ such
that they have normalized 2-point functions {\it
i.e.,}
\eqn\twoponor{\langle O^{I_1} O^{I_2} \rangle = 
{\delta^{I_1I_2} \over |x-y|^{2k}}.}
Their 3-point function is 
\eqn\normthreept{
\langle O^{I_1}(x) O^{I_2}(y) O^{I_3}(z) \rangle ={1\over N}
{\sqrt{k_1 k_2 k_3}\cabc \over
|x-y|^{2\a_3}|y-z|^{2\a_1}|z-x|^{2\a_2}}. 
}
This result is correct only at large $N$ and receives nonzero 
corrections at $O({1\over N^2})$ from non-planar diagrams.

Finally note that the contraction of two or three $\CC's$ may be
related to the integrals of two or three spherical harmonics over the
sphere, by the formulae given in Appendix B.


\newsec{Equations of motion and actions}


\subsec{Foreword to the Calculation}

The particle spectrum of Type IIB SUGRA has been worked
out in \rkim.  The particles are grouped into supermultiplets
\rgunaydin. It turns out that the supermultiplets present in the
theory correspond to representations of the superconformal algebra
labeled by $SU(4)$ weight $(0,k,0)$, $SO(4)$ $j_1=j_2=0$ and scaling
dimension $\e=k$ \rzaf . According to the results of \refs{\rdobrev,
\rminwalla } these are short representations. These supermultiplets of
particles must correspond to CPOs (and their descendents) in \SYMF\
with the same $SO(6)$, $SO(4)$ and scaling dimension labels.  These
are the operators discussed at the beginning of section 2. The $AdS$
fields that correspond to CPOs are particles in the $SU(4)$
representation with weight $(0,k,0)$, $SO(4)$ representation with
$j_1=j_2=0$ and mass $m^2=\e (\e-4)=k(k-4)$ \rwit .  Studying \rkim\
(table III in particular), we conclude that the required fields $s^I$ are
 mixtures of the trace of the graviton on the sphere, and the
five form field strength on the sphere.

Before identifying these fields and starting the calculation we make a
few comments.
\item{1.} 
Since gravity is a gauge theory, not all fields in the
IIB SUGRA action are physical.  We need to choose a gauge and then
solve  the Gauss law constraints to identify the physical fields.
Only these correspond to operators of the \SYMF . 
\item{2.}
Because of the absence of a simple covariant action for IIB SUGRA, 
we choose to work with equations of motion rather than an action. 
In order to compute the action for the fields $s^I$ to cubic order, 
we compute their equations of motion to quadratic order, 
and then produce an action that leads to these equations of motion. 
The action thus produced is of uncertain normalization; we fix this
ambiguity by comparison with the correctly normalized action 
proposed in \rtsorkin, at quadratic order. 
\item{3.}
We need to identify the SUGRA fields $\ph_1, \cdots , \ph_n$ that
couple to various operators only at linear order in fluctuations
about the $AdS_5 \times S^5$ background. Nonlinear higher order
corrections  modify the computed correlation functions of the
corresponding operators only by contact terms. This translates in
spacetime to the fact that we compute only S matrix elements which
are not modified by field redefinitions. We use this freedom to simplify
our analysis.

With the cubic action in hand we then use the procedure of 
\refs{\rvishvanath, \rmathur} to obtain the correlation functions of
interest. 

\subsec{The Setting}

The IIB SUGRA equations of motion of the graviton and the 5-form field
strength are 
\eqn\einstein{R_{mn} = {4\over 4!} F_{mijkl} {F_{n}}^{ijkl},}
\eqn\selfduality{F_{m_1 m_2 m_3 m_4 m_5}={1\over 5!}
\ep_{m_1 m_2 m_3 m_4 m_5 n_1 n_2 n_3 n_4 n_5} F^{n_1 n_2 n_3 n_4 n_5}.}
We use units in which the scale $R_0 = (\l\a'^2)^{1/4}$ of the
$AdS_5$ and $S^5$ is set to be unity. See Appendix A for other
conventions. 

The $AdS_5\times S^5$ background solution is
\eqn\eoffourtwo{\eqalign{
&ds^2 = 
{1\over z^2} (-dx_0^2 + dx_1^2 + dx_2^2 + dx_3^2 + dz^2) + d\Omega_5^2,\cr
&R_{\m\l\n\s}=-(g_{\m\n}g_{\l\s} - g_{\m\s}g_{\l\n});\  \ 
R_{\m\n}=-4 g_{\m\n}; \  \ R_1=-20, \cr
&R_{\a\ga\b\d}=(g_{\a\b}g_{\ga\d} - g_{\a\d}g_{\ga\b}); \  \ 
R_{\a\b}=4g_{\a\b}; \  \ R_2= 20, \cr
&\bar{F}_{{\m_1}{\m_2}{\m_3}{\m_4}{\m_5}}
= \ep_{{\m_1}{\m_2}{\m_3}{\m_4}{\m_5}},\  \
\bar{F}_{{\a_1}{\a_2}{\a_3}{\a_4}{\a_5}}
= \ep_{{\a_1}{\a_2}{\a_3}{\a_4}{\a_5}}.\cr}}
Bulk fields of interest are fluctuations about this background. 
Following \rkim, we set
\eqn\bulkfin{\eqalign{
&G_{mn} = g_{mn} + h_{mn},\cr
&h_{\a\b}=h_{(\a\b)}+{h_2 \over 5} ; \  \ g^{\a\b}h_{(\a\b)}=0, \cr 
&h_{\m\n} = h'_{\m\n} -{h_2 \over 3} g_{\m\n},\  \
h'_{\m\n} = h'_{(\m\n)} + {h' \over 5} g_{\m\n};\  \ 
g^{\m\n}h'_{(\m\n)} = 0,\cr
&F=\bar{F} + \d F,\  \
\d F_{ijklm}=\del_{i}a_{jklm}+ 4\  \ {\rm terms} =
5\del_{[i}a_{jklm]}.\cr}} 

We choose to (almost completely) fix diffeomorphic and 4-form gauge
invariance by choosing the de Donder gauge
$\del^{\a}h_{\a\b}=\del^{\a}h_{\m\a} = \del^\a a_{\a\m_1m_2m_3m_4}=0$.
With this choice the most general expansion of these functions about
the sphere is given by \rkim\ (see Appendix B for information on
spherical harmonics).  For our purposes, it suffices to note that
\eqn\sphdecomp{\eqalign{
{h'}_{\m\n} &= \sum Y^{I} {h'}^I_{\m\n},\cr
h_2 &= \sum Y^{I}h_2^{I},
\cr
a_{\a_1\a_2\a_3\a_4} &= \sum \del^{\a}Y^{I}\ep_{\a\a_1\a_2\a_3\a_4}b^{I},
\cr
a_{\m_1\m_2\m_3\m_4} &= \sum Y^{I}a^{I}_{\m_1\m_2\m_3\m_4}.
}}

\subsec{Linear Constraints and Equations of Motion}

The Einstein and self-duality equations about this background have been 
written out to linear order in \rkim. Of interest to us are the three
constraint equations (E3.2), (E2.2) and (M2.2) in that paper,
\eqn\consta{
\left(\half h'^I-{8\over 15} h_2^I \right) \del_{(\a}\del_{\b)}Y^{I}=0,
} 
\eqn\constb{
\left[
\del_{\m}{h'}^{\m\n I} 
-\del^{\n} \left( h'^I-{8\over 15}h^I_2 + 8 b^I \right)
-{8\over 4!} \ep^{\n\m_1\m_2\m_3\m_4}a^I_{\m_1\m_2\m_3\m_4}
\right]\del_{\a}Y^{I}=0,
}
\eqn\constc{
(a^I_{\m_1\m_2\m_3\m_4}+\ep_{\m_1\m_2\m_3\m_4\m_5}
\del^{\m_5}b^I)\del_{\a}Y^I=0,
}
and the dynamical equations for $b$ and $h_2$ (Eq.(2.31) and (2.32) of \rkim),
\eqn\ema{
\left[
\del_m \del^m b^I + \left( \half {h'}^I - {4\over 3} h_2^I \right)
\right] Y^I = 0,
}
\eqn\emb{
\left[ 
(\del_m \del^m -32) h_2^I + 80 \del_\a \del^\a b^I 
+\del_\a \del^\a \left ( {h'}^I -{16\over 15} h_2^I \right)
\right] Y^I = 0.
}

We are interested in modes with $k\geq2$ only. For such modes the constraint
\consta\ may be used to eliminate ${h'}^I$ from \ema\ and \emb\ to yield
\eqn\emc{
\del_{m}\del^{m} b - {4\over 5}  h_2 = 0,
}
\eqn\emd{
(\del_{m}\del^{m} - 32) h_2 + 80 \del_{\a}\del^{\a}b = 0.
}
These two equations may now be diagonalized.  Using the fact that 
$\del_{\a}\del^{\a}Y^I=-k(k+4)Y^I$ as shown in Appendix B,
we find that the diagonal linear combinations  (We choose the
normalization such that the inverse relations are simple:  
$h_2^I=10ks^I + 10(k+4)t^I,\  \ b^I=-s^I+t^I$.),
\eqn\diag{\eqalign{
s^I &={1\over 20 (k+2)}[h_2^I-10(k+4)b^I],\cr
t^I &={1\over 20 (k+2)}[h_2^I+10k{b^I}] \cr 
}}
obey the equations of motion
\eqn\emef{\eqalign{
\del_{\m}\del^{\m}s^I &= k(k-4) s^I, \cr
\del_{\m}\del^{\m}t^I &= (k+4)(k+8) t^I. \cr
}} 

To linear order, $s^I$ corresponds to CPOs in \SYMF, and it will be
the focus of our attention through the rest of the paper.

The scalars
$t^I$, on the other hand, correspond to descendents of CPOs; specifically
they map to the operator $\phi^{(6)}$ in Table 1 of \rgunaydin . 
The expansion of $t$ proportional to the $k^{th}$ spherical harmonic, 
corresponds
to an operator formed by acting with 4 $Qs$ and 4 $\bar{Q}s$ 
on the trace of $k+4$ $\ph$ operators. The 3-point functions of these operators
are determined in terms of those of CPOs by the supercoformal algebra, and 
so we will not compute them directly. Henceforth we set  $t^{I}=0$.

We now construct an action whose variations leads to the equations
of motion of $s^I$.
\eqn\actna{
S = \int \sum {A_I \over 2} [ - (\del_{\m}s^I)^2  - k(k-4)  (s^I)^2 ] 
}
with $A_I$ undetermined constants which depend on $k$.

\subsec{Normalization of the Quadratic Action}

The normalization coefficients $A_I$ may be determined by comparison of 
\actna\ with the full `actual' action of IIB SUGRA \rtsorkin\
\eqn\sorokin{
S = {1\over 2\kappa^2} \bar{S}
= {1\over 2\kappa^2} \int d^{10}x \sqrt{-G} 
\left\{R - {8 \over 4!} {\del^m a \del_n a \over (\del a)^2} 
(F-\tilde{F})_{mijkl} \tilde{F}^{nijkl} \right\},
}
where $\tilde{F}$ is defined by the right-hand-side (RHS) of \selfduality, 
and $a$ is an auxiliary field.
In our units ${1 \over 2\kappa^2} = {4N^2\over (2\pi)^5}$.

In order to obtain $A_I$ from \sorokin\ we work at
quadratic order, choose a gauge, solve for all constrained fields in
terms of physical fields, and then set all physical fields except
$s^{I}$ to zero.  

Firstly we eliminate the auxiliary field $a$
in \sorokin .  As shown in \rtsorkin, we are free to fix a gauge by
choosing an arbitrary function for $a$. We will set $a = x^4$, which
amounts to removing the components of the 4-form potential of the form
$A_{ijk4}$.

Having done this use \consta, \constb, (but not yet \constc ) 
in \sorokin\ and set all unconstrained 
fields other than $b$ and $h_2$ and $h'_{\m\n}$ to zero. 
The action we obtain at the end
of this process is ($z(k)$ is defined in Appendix B equation (B.4))
\eqn\fivedaction{
\bar{S} = \sum_{I} z(k) \int d^5x \sqrt{-g_1} 
\left\{ L^I_1 + L^I_2 + L^I_3 \right\}.
}
$L^I_1$ contains terms from the Einstein part of the action except those
 involving ${h'}^I_{(\m\n)}$,
\eqn\llone{
L^I_1 = -{2\over 15} \left\{ (\del h_2^I)^2 + k(k+4)(h_2^I)^2 \right\}
+ {32 \over 1875} \left\{ 8(\del h_2^I)^2 +5(k^2+4k-9)(h_2^I)^2 \right\},
}
where the first group comes from the $h_2$ kinetic and mass terms
while the second group was obtained by inserting \consta\ into the
$h'$ kinetic and mass terms.  $L^I_2$ contains terms from the $F^2$
part except ${h'}^I_{(\m\n)}$ terms,
\eqn\lltwo{
L^I_2 = -8k(k+4) \left\{
(\del b^I)^2 + k(k+4) (b^I)^2 + {8\over5}h_2^I b^I \right\} - {352\over
125}(h_2^I)^2 
}
$L^I_3$ is the part of \sorokin\ quadratic in ${h'}^I_{(\m\n)}$ :
\eqn\llthree{\eqalign{
L^I_3=
&-{1\over 4}\del_\l {h'}^I_{(\m\n)}\del^\l {h'}^{I(\m\n)}
+\half \del^{\m}{h'}^I_{(\m\n)}\del_{\l}{h'}^{I(\l\n)}
-{8\over 25} \del^{\m}h_2^I \del^{\n} {h'}^I_{(\m\n)}
\cr
& -{1\over 4} (k^2+4k-2) {h'}^I_{(\m\n)} {h'}^{I(\m\n)}.
}}

We now attempt to use $\constb $ to obtain the quadratic dependence of
$L^I_3$ on $b$ and $h_2$.  On eliminating ${h'}^I$ and
$a^I_{\m_1\m_2\m_3\m_4}$ from \constb\ and separating out the trace
explicitly we obtain
\eqn\ccb{
\del^{\m}{h'}^I_{(\m\n)} 
= \del_{\n}\left\{ {8\over 25} h_2^I+16 b^I \right\}.
}
We can solve the equation by setting 
$${h'}^I_{(\m\n)}=H^I_{(\m\n)} + \del_{(\m}\del_{\n)}K^I,$$ 
where $H^I_{(\m\n)}$ obeys $\del^{\mu}H_{\mu\nu}=0$ and $K^I$ satisfy
$(\del^2 -5)K^I = {2\over 5}h^I_2 + 20b^I$.  Note that unlike
$h_{\mu\nu}$, $H$ may consistently be set to zero for arbitrary $h_2$
and $b$.  Substituting this into $L^I_3$ leads unfortunately to an
action non-local in $b$ and $h_2$.

To avoid undue complications, we notice that it is sufficient for us
to compute \fivedaction\ on shell in order to obtain $A^I$. 
In that case
$$K^I = {2\over 5(k+1)(k+3)} (h^I_2 - 30b^I)$$
We substitute $h_2^I = 10ks^I$, $b^I = -s^I$ in \actna\ to find
\eqn\actionss{\eqalign{
L[s^I] = 
&- \left\{ {64\over 5}k^2 + 32k -128 \right\} (\del s^I)^2
-{32 \over 5} k^2(2k+1)(k-4)(s^I)^2
\cr
&-{4(k^2+4k-2) \over (k+1)^2}(\del_{(\m}\del_{\n)}s^I)^2
-{4 \over (k+1)^2} (\del_\l\del_{(\m}\del_{\n)}s^I)^2.
}}
\fivedaction\ vanishes on shell in the bulk (as every quadratic action
does), but is nonzero as a function of boundary values due to surface
terms. We now compute each of \actna\ and \fivedaction\ as a function
of boundary values of $s^I$, and compare the two results to read off
the value of $A^I$. The result is
\eqn\quadnorm{
A_I = 32 { k(k-1)(k+2) \over k+1} z(k).
}

\subsec{Cubic Couplings}

To study the 3-point functions of the field $s^{I}$, we need the cubic
terms in the action \actna.  To compute these we need quadratic
corrections to Eqs. \consta, \constc, \ema\ and \emb. We define
\eqn\constquad{\eqalign{
{h'} &= {16\over 15} h_2 + 10 Q_1,
\cr
a_{\m_1\m_2\m_3\m_4} &= - \ep_{\m_1\m_2\m_3\m_4\m_5} 
(\del^{\m_5} b + Q_3^{\m_5}),
}}
\eqn\eomqa{\eqalign{
(\del_m\del^m -32) h_2 + 80 \del_\a\del^\a b 
+ \del_\a\del^\a (h' - {16\over 15} h_2) = 10 Q_2,
\cr
5\del_{[\m_1} a_{\m_2\m_3\m_4\m_5]} = \ep_{\m_1\m_2\m_3\m_4\m_5} 
\left[ \del_\a\del^\a b + \half h' - {4\over 3} h_2 + Q_4 \right].
}}
Substituting \constquad\ into \eomqa, we obtain,
\eqn\eomqb{\eqalign{
(\del_m\del^m-32)h_2 + 80\del_\a\del^\a b + 10(\del_\a\del^\a Q_1 -
Q_2) &= 0, 
\cr
\del_m\del^m b - {4\over 5} h_2 + 5Q_1 + \del_\m Q_3^\m + Q_4 &= 0.
}}
The corrected equation of motion for $s$ is a linear combination of
the two above: 
\eqn\eomqs{
[\del_\m\del^\m - k(k-4)]s^I = {1\over 2(k+2)} 
\left\{
(k+4)(k+5)Q_1 + Q_2 + (k+4) (\del_\m Q_3^\m + Q_4)\right\}^I.
}

To calculate the $Q_i's$ we use the methods outlined in \rkim.  The
first lines of \constquad\ and \eomqa\ are the coefficients of
$\del_{(\a}\del_{\b)}Y^I$ and $Y^I g_{\a\b}$ respectively, in the
equation $R_{\a\b}={4\over
4!}F_{\a\a_1\a_2\a_3\a_4}F_{\b}^{\a_1\a_2\a_3\a_4}$.  To compute $Q_1$
and $Q_2$, we must therefore compute $R_{mn}$ and $
F_{mijkl}{F_n}^{ijkl}$ to second order in $s$ \rmtw.  Since we are
only interested in the $s$ dependence of these quantities, we
substitute
\eqn\subst{\eqalign{
h^I_{\m\n} &= -{3\over 25} h^I_2 g_{\m\n} 
+ {2\over 5(k+1)(k+3)} \del_{(\m}\del_{\n)} (h^I_2 - 30b^I) \cr
&= U(k) s^I g_{\m\n} + W(k) \del_{(\m}\del_{\n)}s^I,
\cr
h^I_{\a\b} &= {h^I_2\over 5}g_{\a\b} = V(k) s^I g_{\a\b}, \cr
b^I &=  X(k) s^I,\       \ h_{\a\m} = 0,
\cr
}} 
\eqn\defns{
V(k) = -{5 \over 3} U(k) = 2 k,\  \ 
W(k) = {4 \over (k+1)},\  \ X(k) = -1.
}   
to find
\eqn\eomquad{\eqalign{
R_{\a\b} = &\half (Y  + {1 \over 10} Z_{\gamma}^{\gamma})g_{\a\b} + 
{1\over 4} Z_{(\a\b)},
\cr
Y \equiv & V_1V_2\del^\ga (s_1\del_\ga s_2) 
+ U_1 V_2\del^\m (s_1\del_\m s_2)
+ W_1 V_2  \del_\m(\del^{(\m}\del^{\n)} s_1\del_\n s_2),
\cr
Z_{\a\b} \equiv & (3V_1V_2 + 5 U_1U_2)
( \del_\a s_1 \del_\b s_2 + 2 s_1 \del_\a \del_\b s_2 ),
\cr
&+W_1W_2 ( \del_\a\del^{(\m}\del^{\n)} s_1 \del_\b
\del_{(\m}\del_{\n)} s_2  
+ 2\del^{(\m}\del^{\n)} s_1 \del_\a \del_\b \del_{(\m}\del_{\n)} s_2 )
\cr
{4\over 4!} F_{\a ijkl}{F_\b}^{ijkl} = 
& 4 g_{\a\b} \{ X_1X_2 (\del^\ga\del_\ga s_1 \del^\d\del_\d s_2
+\del^\m\del^\ga s_1 \del_\m\del_\ga s_2 )
\cr
&-8 V_1X_2 s_1 \del^\ga \del_\ga s_2 + 10 V_1V_2 s_1 s_2 \}- 8 X_1X_2
\del_\a \del_\m s_1 \del_\b \del^\m s_2.
}}
In the equations above, the symbol $s_i$ is used as shorthand for 
$s^{I_i} Y^{I_i}$ and $U_i,\cdots, X_i$ as shorthand for 
$U(k_i), \cdots, X(k_i)$, respectively. 
Summation over $I_1$ and $I_2$ is assumed.

Projection of these quantities onto  $\del_{(\a}\del_{\b)}Y^{I}$
yields 
\eqn\qone{\eqalign{
Q_1^{I_1} &= {1\over 20 q(k_1) z(k_1)} \sum_{2,3} \left\{
(c_{123} + d_{231} + d_{321} )T^{23} 
+ 32 X_2 X_3 c_{123} \del_\m s_2 \del^\m s_3 \right\},
\cr
T_{23} &\equiv (3V_2V_3 + 5U_2U_3)s_2s_3
+ W_2W_3 \del^{(\m}\del^{\n)} s_2 \del_{(\m}\del_{\n)} s_3, 
}}
where $c_{123}$, {\it etc.,} are used as shorthand for
$c(k_1,k_2,k_3)$,  {\it etc.} defined in Appendix B (dropping an
overall factor of $\cabc$ from the equations which will be reinstate
later) and $s_i$ as shorthand for $s^{I_i}$.
 
Projection onto $g_{\a\b}Y^{I}$ yields
\eqn\qtwo{\eqalign{
Q_2^{I_1} &= {1\over 20z(k_1)} \sum_{2,3} \left\{
10 S_{123} +  T_{23} (b_{123} - 2 f_3 a_{123}) 
+ 32 X_2 X_3 \del_\m s_2 \del^\m s_3 b_{123} \right\},
\cr
S_{123} &\equiv 
- V_2V_3 b_{213} s_2 s_3 + V_3U_2 a_{123} \del^\m(s_2\del_\m s_3)
+ W_2V_3 a_{123} \del_\m(\del^{(\m}\del^{\n)} s_2\del_\n s_3) 
\cr
&-8X_2X_3 ( a_{123}f_2f_3s_2s_3 + b_{123} \del^\m s_2 \del_\m s_3 ) 
-a_{123} ( 64V_2X_3 f_3 + 80 V_2 V_3 ) s_2 s_3.
}}

Expansion of the self-duality equations to quadratic order and
projection onto appropriate spherical harmonics yields $Q_3$ and
$Q_4$. $Q_3$ arises as the coefficient of $\del_\a Y^{I}$ and $Q_4$
from the coefficient of $Y^{I}$ in the self duality equation
\selfduality.  The answers are
\eqn\qthree{
Q_3^{\m I_{1}} = - {1\over f(k_1)z(k_1)} \sum_{2,3} 
\left\{
(U_2 +3V_2)X_3 s_2 \del^\m s_3 
+ W_2 X_3 \del^{(\m}\del^{\n)} s_2 \del_\n s_3 
\right\} b_{213},
}
\eqn\qfour{
Q_4^{I_1} = - {1\over 4z(k_1)} \sum_{2,3} \left\{
T^{23} - (16V_2X_3f_3 + 40 V_2V_3) s_2s_3 \right\} a_{123},
}
where $f(k)\equiv k(k+4)$ and $T^{23}$ is the same as in \qone.

This completes the evaluation of the RHS of the equation
of motion \eomqs\ which now takes the form
\eqn\qdstr{
(\del_\m \del^\m - m_{I_1}^2)s^{I_1} = \sum_{I_2,I_3} \left\{
D_{I_1I_2I_3} s^{I_2}s^{I_3} 
+ E_{I_1I_2I_3} \del_\m s^{I_2} \del^\m s^{I_3} 
+ F_{I_1I_2I_3} \del^{(\m}\del^{\n)} s^{I_2}
\del_{(\m}\del_{\n)}s^{I_3} \right\}.
}
where $D$, $E$ and $F$ are computed by substituting \qone, \qtwo,
\qthree\ and  \qfour\  into \eomqs.  We can remove the derivative
terms on the RHS of \qdstr\ by a field redefinition 
\eqn\redef{
s^{I_1} = s'^{I_1} + \sum_{I_2, I_3} \left\{
J_{I_1I_2I_3}s'^{I_2}s'^{I_3} + L_{I_1I_2I_3}\del^\m s'^{I_2} \del_\m
s'^{I_3} \right\},
}
where
$$L_{I_1I_2I_3} = {1 \over 2} F_{I_1I_2I_3},\  \
J_{I_1I_2I_3} = \half E_{I_1I_2I_3} 
+ {1 \over 4} F_{I_1I_2I_3} (m_{I_1}^2 - m_{I_2}^2 - m_{I_3}^2 + 8), $$
such that \qdstr\ becomes 
(we henceforth omit the primes on redefined fields)
\eqn\eomf{(\del_\m \del^\m - m_{I_1}^2) s^{I_1} = 
\sum_{I_2,I_3} \lambda_{I_1I_2I_3} s^{I_2} s^{I_3},
}
where
\eqn\dprime{
\lambda_{I_1I_2I_3} = 
D_{I_1I_2I_3} - (m_{I_2}^2 + m_{I_3}^2 - m_{I_1}^2)J_{I_1I_2I_3}
- {2 \over 5} L_{I_1I_2I_3} m_{I_2}^2 m_{I_3}^2.
}
 
Putting together the values of $Q_i's$ and reintroducing the factors
of $\cabc$  
that we suppressed for notational convenience (the definition of $\Sigma$ 
and $\a_{i}$ are as in Section 2) we obtain
\eqn\ans{\eqalign{ 
\lambda_{I_1I_2I_3} = { a(k_1,k_2,k_3) \over 2 z(k_1) (k_1 +2)}  
{8 \Sigma \{ (\half\Sigma)^2-1\} \{(\half\Sigma)^2-4 \} \a_1 \a_2 \a_3
\over (k_1 - 1)k_1(k_2+1)(k_3+1)} \cabc.
}} 
Taking into account the normalization of the quadratic action \quadnorm,
the cubic coupling constant is 
\eqn\thpt{\eqalign{
\CG_{I_1I_2I_3} 
&= A_{I_1} \lambda_{I_1I_2I_3}
\cr
&=  a(k_1,k_2,k_3) 
{ 128 \Sigma \{(\half\Sigma)^2-1\} \{(\half\Sigma)^2-4\} \a_1\a_2\a_3
\over (k_1 +1)(k_2+1)(k_3+1)} \cabc.
}}

Note that $\CG_{I_1I_2I_3}$ is totally symmetric, which
ensures that the equations of motion can be derived from an action.


\newsec{The strong coupling limit of the three-point function}


The cubic equations of motion for the fields $s^I$ may be derived from 
the action 
\eqn\actiona{
S={4 N^2\over (2\pi )^5}\int d^5x \sqrt{-g_1} \left\{ 
\sum_{I} {A_I\over 2}[-(\del s_I)^2 - k(k-4) (s_I)^2]
+\sum_{I_1,I_2,I_3}{1 \over 3}\CG_{I_1I_2I_3}s^{I_1}s^{I_2}s^{I_3}
\right\}.
} 
There is an ambiguity in the use of this action due to our lack of
knowledge of $S_{int}$. We know only that the field that couples to
the primary operator of interest is proportional to $s^I$. The unknown
proportionality constant may be a function of $N$, $\l$ and $k$.  Let
$\tilde{s}^I$ be the field that couples to CPOs via $S_{int} = \int
\tilde{s}^I \CO^I$ and $s^I=w^I\tilde{s}^I$ for some function $w^I$.
The action written in terms of $\tilde{s}$ is
\eqn\actionbb{\eqalign{
S=&{4 N^2\over (2\pi)^5}\int d^5x \sqrt{-g_1}\cr 
& \left\{
- \sum_{I} {A_I (w^I)^2\over 2}[(\del \tilde{s}^I )^2 +
k(k-4)(\tilde{s}^I)^2] 
 + \sum_{I_1I_2I_3}{w^{I_1}w^{I_2}w^{I_3} \CG_{I_1I_2I_3} \over 3}
\tilde{s}^{I_1} \tilde{s}^{I_2} \tilde{s}^{I_3}
\right\}. \cr}} 

To compute the 2- and 3-point functions of CPOs from \actionbb\ we
apply the formulae derived, for instance, in \rmathur. From Eq. 17 and
the correction factor in Eq. 95 of \rmathur, we derive that in the
large $N$ limit of \SYMF,
\eqn\twopta{
\langle \CO^{I_1}(x) \CO^{I_2}(y) \rangle = 
{4 N^2 \over (2\pi)^5}
{1\over \pi^2} {\G(k+1)\over \G(k-2)}{ 2k-4\over k}
{ A_I (w^I)^2 \delta^{I_1I_2} \over |x-y|^{2k}}.
}
{}From Eq. 25 of the same paper we derive that
\eqn\threepta{
\langle \CO^{I_1}(x) \CO^{I_2}(y) \CO^{I_3}(z) \rangle = 
-{4 N^2 \over (2\pi)^5}
{1\over \pi^4}
{\G(\a_1)\G(\a_2)\G(\a_3) \G(\half \Sigma -2) 
\over \G(k_1 -2)\G(k_2 -2 )\G(k_3 -2 )} 
{w^{I_1}w^{I_2}w^{I_3}\CG_{I_1I_2I_3} \over 
|x-y|^{\a_3} |y-z|^{\a_1} |z-x|^{\a_2} }.
}
Using \quadnorm\ and the formula (B.1), we can simplify \twopta\ to
read 
\eqn\twoptb{
\langle \CO^{I_1}(x) \CO^{I_2}(y) \rangle = 
{4 N^2 \over (2\pi)^5} {\pi \over 2^{k-7}}
{ k (k-1)^2 (k-2)^2 \over (k+1)^2 } 
{ (w^I)^2 \delta^{I_1I_2} \over |x-y|^{2k}}.
}
Similarly using \thpt\ and (B.2) we have, 
\eqn\threeptb{\eqalign{
\langle \CO^{I_1}(x) \CO^{I_2}(y) \CO^{I_3}(z) \rangle
= &- {4 N^2 \over (2\pi)^5} 
{1 \over \pi 2^{\half \Sigma -9}}
{ w^{I_1}w^{I_2}w^{I_3} \cabc \over |x-y|^{\a_1} |y-z|^{\a_2} |z-x|^{\a_3}} 
\cr 
&\times \left\{
{k_1(k_1-1)(k_1-2)\over (k_1+1)} {k_2(k_2-1)(k_2-2)\over (k_2+1)}
{k_3(k_3-1)(k_3-2)\over (k_3+1)}
\right\}.
}}
Finally, we obtain the 3-point functions of normalized CPOs,
\eqn\normthreeptb{
\langle O^{I_1}(x) O^{I_2}(y) O^{I_3}(z) \rangle 
= {1 \over N} 
{\sqrt{k_1k_2k_3}\cabc \over |x-y|^{2\a_3} |y-z|^{2\a_1}|z-x|^{2\a_2}},
}
which agree exactly with the weak coupling result \normthreept\ in
Section 2.  Note that all the numerical factors as well as the unknown
function $w^I$ present in \twoptb\ and \threeptb\ have been canceled.

The action \actionbb\ was obtained up to an overall normalization merely 
from the equations of motion. To obtain this overall normalization, 
we had to make assumptions about the `true' action (including surface terms) 
for IIB SUGRA. Changing the normalization of \actionbb\ by a factor
$\eta $ scales the result in \normthreeptb\ by ${1\over \sqrt{\eta}}$, 
i.e a factor independent of $k$.
We present here a further argument that the 3-point functions in 
\normthreeptb\ are correctly normalized. 

$R$-symmetry currents are descendents of $\Tr(\ph^i\ph^j)$ (after
subtracting the trace).  Specifically
$J^b_{\a\dot \b a}\propto \ep^{ijk(b}(Q_{\a i }\bar{Q}_{\dot \b}^l - {1
\over 4} Q_{\a m} \bar{Q}_{\dot \b}^m \delta_i^l)
{\bf Tr}(\ph_{a)j}\ph_{kl} - {1 \over 4!} \ph_{mn} \ph_{pq}\ep^{mnpq}
\epsilon_{a)jkl})$.    Here $a, b,
\cdots = 1, \cdots , 4$ label  the {\bf 4} or
${\bf \bar{4}}$ of $SU(4)$, brackets indicate trace removal, 
 $\a$ is a chiral spinor index and $\dot
\b$ is an anti-chiral spinor index.  Therefore, the correlation
functions of $R$-symmetry currents are determined in terms of those of
$\Tr(\ph^i\ph^j)$.  However, the 2- and 3-point functions of
$R$-symmetry currents are known to be given exactly by the free field
value (\rmathur , and references cited therein ). This is sufficient
to ensure that the overall normalization in \normthreeptb\ agrees with
the free field result at least for $k=2$, and hence for all $k$.

\bigskip
\centerline{\bf Acknowledgment}
\medskip

We are grateful to Tom Banks, Micha Berkooz, Sumit Das, Rajesh Gopakumar, 
Michael Green, Igor Klebanov, Andrei Mikhailov, K. Narayan and 
Edward Witten for helpful
discussions.  The work of S.L. and S.M. was supported in part by DOE grant
\#DE-FG02-91ER40671, M.R. by a Princeton University Fellowship, 
and N.S. by DOE grant \#DE-FG02-90ER40542.
 

\appendix{A}{Notations and Conventions}

Consider the manifold $AdS_5 \times S^5$.  We use Latin indices
$i,j,k,l,m,\ldots$ for the whole $10$-dimensional manifold.  Indices
$\m,\n,\l, \ldots$ are $AdS_5$ indices and run from $0$ to $4$.
Indices $\a,\b,\ga, \ldots$ are $S^5$ indices and run from $5$ to $9$.
Our choice of the signature of the metric is $(-+\ldots +)$.

We use $G_{mn}$ for the metric and $g_{mn}$ for its background value. 
The conventions for metric connection, curvature tensor, 
Ricci tensor and the scalar curvature are
\eqn\christofetc{\eqalign{
\G^i_{jk} &= \half G^{il} ( \p_k G_{lj} + \p_j G_{lk} - \p_l G_{jk}),
\cr
{R^i}_{jkl} &= 
\p_k \G^i_{jl} - \p_l \G^i_{jk} + \G^i_{km} \G^m_{lj} - \G^i_{lm}
\G^m_{kj}, \cr
R_{jl} &= {R^i}_{jil}, \  \ R = G^{jl} R_{jl}.
}}
For comparison, we note that Ref. \rkim\ uses the same convention as
ours except that they define $R_{mn} = R^k_{mnk} = - R^{k}_{mkn}$.

The determinant of the metric is denoted by $G$.  The determinants of
the $AdS$ metric $g_{\m\n}$ and the $S^5$ metric $g_{\a\b}$ are
denoted by $G_1$ and $G_2$, respectively.  The completely
antisymmetric $\ep_{m_0 \cdots m_9}$ symbol is defined to be a tensor
of rank 10, such that $\ep_{0123456789} = \sqrt{-G}$ and
$\ep^{0123456789} = - 1/ \sqrt{-G}$.


\appendix{B}{Spherical Harmonics}

\subsec{Scalar Spherical Harmonics}

The set of scalar functions on $S^{D}$ form a 
vector space which is an infinite dimensional reducible
representation of $SO(D+1)$. Scalar spherical harmonics (SSH) form 
a complete basis on this space.

It is convenient to regard a function on $S^{D}$ as a restriction of
functions on the $R^{D+1}$ in which $S^{D}$ is embedded. An arbitrary
$C^{\infty}$ function on $R^{D+1}$ may be expanded in polynomials in
the Cartesian coordinates $x^i$, so it is sufficient to consider
separately functions on $R^{D+1}$ homogeneous in $x^i$ of degree $k$.
Not all such functions are independent when restricted to a sphere.
Consider, for example, $r^2x^{i_1}...x^{i_k}$. This is a function of
degree $k+2$ but when restricted to the sphere, it is identical to
$x^{i_1}...x^{i_k}$, a function of degree $k$.  If at each degree we
wish to restrict ourselves to functions linearly independent of those
of lower degree, we must consider only functions
$\CC_{i_1...i_k}x^{i_1}...x^{i_k}$ such that
$\CC_{i_1...i_k}\d^{i_mi_n}=0$ for any $1 \le m,n \le k$.  With no
loss of generality, we may demand that $C_{i_1..i_k}$ be symmetric in
$i_1..i_k$.

Thus we have shown that each independent component of a totally
symmetric traceless tensor of rank $k$ defines a spherical harmonic by
$Y^I = \CC^I_{i_1...i_k}x^{i_1}...x^{i_k}$.  This construction clearly
shows which representation of $SO(D+1)$ the spherical harmonics
transform in.  The Gelfand-Zetlin indices for the representation are
$(h_1, h_2, h_3, \ldots)=(k,0,0, \ldots)$.  The degeneracy of the
harmonics is the number of symmetric polynomials of degree $k$ minus
the number of symmetric polynomials of degree $k-2$, {\it i.e.}
${D+k\choose D}-{D+k-2\choose D}$.

Since $M_{\a\b}=(-i) (x_{\a}\del_{\b}-x_{\b}\del_{\a})$, the Casimir
of this representation, $L^2 = \half M_{\a\b}M^{\a\b}=k(k+D-1)$, is
simply the value of $-r^2\del^2$ on the sphere. Therefore we deduce
that the degree $k$ spherical harmonics are eigenvectors of $\del^2$
on the sphere, with eigenvalues $-k(k+D-1)$.

The eigenvalue of $\del^2$ may be obtained in an alternative
fashion. Note that the harmonics as polynomials in $R^{D+1}$ obey
$(\del^2)_{D+1} f_{k}=0$. However, $(\del^2)_{D+1}={1\over r^{D}}\p_r
r^{D}\p_r +{1\over r^2}(\del^2)_{S^D}$.  Since our functions behave as
$r^{k}$, $(\del^2)_{D+1} f_k=0$ implies $(\del^2)_{S^D}
f_k=-k(k+D-1)f_k$ as above.

\subsec{Scalar Harmonic Contractions on $S^5$}

We need to evaluate the integral of the product of two or three scalar
spherical harmonics over $S^5$.  Let $Y^{I} =
\CC^I_{i_1...i_k}x^{i_1}...x^{i_k}$ be the spherical harmonics.  The
results of the integration are
\eqn\twoint{
{1\over \omega_5} \int_{S^5} Y^{I_1} Y^{I_2} 
= {\delta^{I_1I_2} \over 2^{k-1} (k+1)(k+2) }, 
}
\eqn\threeint{
{1\over \omega_5} \int_{S^5} Y^{I_1} Y^{I_2} Y^{I_3}
= {1 \over (\half \Sigma +2)! 2^{\half(\Sigma -2)}}
{k_1!k_2!k_3! \over \a_1! \a_2! \a_3!} \cabc,
}
where $\a_1=\half(k_2+k_3-k_1)$, $\Sigma=k_1+k_2+k_3$ and 
$\omega_5=\pi^3$ is the area of a unit 5-sphere. 
Both of the above equations can be derived 
by using the following general formula:
\eqn\wick{
{1\over \omega_5} \int_{S^5} x^{i_1} \cdots x^{i_{2m}} = 
{2^{1-m} \over (m+2)!} \times ({\rm All\  \ possible\  \ contractions}),
}
where ``All possible contractions'' means $\delta^{i_1i_2}$ for $m=1$,
$\delta^{i_1i_2}\delta^{i_3i_4} + \delta^{i_1i_3}\delta^{i_2i_4} +
\delta^{i_1i_4}\delta^{i_2i_3}$ for $m=2$ and analogous objects for
higher $m$. This formula can be proved by starting from
$$ \int_{S^5} x^{i_1} \cdots x^{i_{2m}} =
{\p^{2m} \over \p J_{i_1} \cdots \p J_{i_{2m}}} \int_{S^5} e^{J\cdot x}.$$

The following integrals occur naturally 
when one considers projection of the equations in section 4 
onto appropriate spherical harmonics.  
\eqn\sphcont{\eqalign{
\int Y^{I_1} Y^{I_2} &= z(k) \delta^{I_1I_2},
\cr
\int \del_{(\a}\del_{\b)}Y^{I_1} \del^{(\a}\del^{\b)}Y^{I_2} &= 
q(k)z(k) \delta^{I_1I_2},
\cr
\int Y^{I_1} Y^{I_2} Y^{I_3} 		&= a(k_1,k_2,k_3) \cabc,
\cr
\int Y^{I_1} \del_{\a}Y^{I_2} \del^{\a}Y^{I_3} &= b(k_1,k_2,k_3) \cabc,
\cr
\int \del^{(\a}\del^{\b)}Y^{I_1} \del_{\a}Y^{I_2} \del_{\b}Y^{I_3} 
&= c(k_1,k_2,k_3) \cabc,
\cr
\int Y^{I_1} \del^{(\a}\del^{\b)}Y^{I_2} \del_{\a}\del_{\b}Y^{I_3} 
&= d(k_1,k_2,k_3) \cabc.
}}
The functions $q, a, b, c, d$ can be evaluated by integrating by parts
and using the fact that $\del_\a \del^\a Y^I = -k(k+4) Y^I$.

\subsec{Vector and Tensor Spherical Harmonics}

One may now ask for a basis in the space of vector functions on the
sphere. To find such a basis, one again considers vectors of the form
$e_a Y^a = e_a C^a_{i_1...i_k}x^{i_1}...x^{i_k}$ in $R^{D+1}$, where
$e_a$ is a unit vector in the $a^{th}$ Cartesian direction.  This is a
complete set of vector functions on $R^{D+1}$ but is over-complete on
$S^{D}$ for two reasons.  The first is the same as that for SSH and
may be fixed in the same fashion.  The second reason is that some of
these vectors have no projection onto the tangent space of the sphere.

The vector function $e_a C^a_{i_1...i_k}x^{i_1}...x^{i_k}$ transforms
in the product of the vector representation and $(k,0,\ldots ,0)$
under $SO(D+1)$.  For the rest of this subsection we assume that $D$
is odd as it is in our paper.  That product has 3 irreducible
representations, $(k-1,0,0,\ldots,0)$, $(k+1,0,0,\ldots,0)$ and
$(k,1,0,\ldots,0)$.  The first corresponds to a vector of the form
$Y^a=x^a Y(k-1)$, where $Y(k-1)$ is a SSH of degree $k-1$.  It has no
projection onto the tangent space of $S^{D}$.  The second corresponds
to a vector of the form $Y^a=\p^a Y(k+1)$.  It is a derivative of a
SSH of one higher degree.  Projected onto $S^{D}$, this becomes
$\del_{\a}Y(k+1)$.  The last corresponds to vector functions that obey
$x_a Y^a=\p_a Y^a=0$, which implies $\del_{\a}Y^{\a}=0$ on $S^D$.
This function is called a vector spherical harmonic.

In summary, an arbitrary vector function on $S^{D}$ is a linear
combination of the gradients of SSH and vector spherical harmonics
introduced above.

The story is very similar for symmetric tensors.   
Any symmetric tensor on the sphere can be decomposed into a sum of the form
$$S_{\a\b}=\sum_{I} g_{\a\b} A^I Y^I+
B^I\del_{(\a}\del_{\b)}Y^I +C^I\del_{(\a}Y_{\b)}^I+
D^IY_{(\a\b)}^I,$$
where $A,B,C,D$ are constants.  The $Y_{\b}^I$ and $Y_{(\a\b)}^I$ are
vector and symmetric tensor spherical harmonics.  Symmetric tensor
spherical harmonics of degree $k$ are a new set of functions.  They
transform in the $(k,1,1,0,..0)$ representation of $SO(D+1)$ and obey
$$\del^{\b} Y^I_{\b}=\del^{\b}Y_{(\a\b)}^I=g^{\a\b}Y_{(\a\b)}^I=0$$
These properties, and the orthogonality of SSHs on the sphere imply that 
$$A^I= {1\over D}
{\int S_{\a\b}g^{\a\b} Y^I\over 
\int Y^I Y^I };\      \ 
B^I= {\int S_{\a\b}\del^{(\a}\del^{\b)}Y^I\over 
\int \del_{(\a}\del_{\b)}Y^I \del^{(\a}\del^{\b)} Y^I}$$

\listrefs

\end